\documentclass[journal,twoside,web]{ieeecolor}
\usepackage{generic}
\usepackage{cite}
\usepackage{amsmath,amssymb,amsfonts}
\usepackage{algorithmic}
\usepackage{graphicx}
\usepackage{algorithm,algorithmic}
\usepackage{hyperref}
\usepackage{CJKutf8}

\usepackage{textcomp}
\usepackage{mathtools}
\usepackage{bm}
\usepackage{tabularx}
\usepackage{mathrsfs}
\usepackage{comment}

\usepackage{graphics}
\usepackage{here}
\def\BibTeX{{\rm B\kern-.05em{\sc i\kern-.025em b}\kern-.08em
    T\kern-.1667em\lower.7ex\hbox{E}\kern-.125emX}}
\begin{document}
\title{Impact of Multiple Non-Invasive Biosignals on Cardiovascular Biomarker Estimation via Simulation-Based Inference}
\author{Shusaku Maeda, Masahiro Nakano, Tomoharu Iwata, Kenji Komiya, Ryo Nishikimi
 and Kunio Kashino
\thanks{Shusaku Maeda, Masahiro Nakano,
 Tomoharu Iwata, Kenji Komiya, Ryo Nishikimi and Kunio Kashino are with Communication Science Laboratories, NTT Corporation, Atsugi, Kanagawa 243-0198, Japan (e-mail: shusaku.maeda@ntt.com).}}
\maketitle
\thispagestyle{empty}
\begin{abstract}
As the population ages, the number of patients with cardiovascular diseases continues to increase, highlighting the need for early detection before progression to severe and irreversible functional decline. Consequently, estimating cardiovascular biomarkers from non-invasive biosignals, such as photoplethysmography (PPG) and arterial pressure wave (APW) signals, has attracted increasing attention. These signals can be measured using wearable and cuff-type devices. Previous studies have used PPG and APW signals to estimate cardiovascular biomarkers. However, these signals exhibit strong similarities in both the temporal and frequency domains and primarily reflect peripheral and arterial pulse waveforms. Therefore, they may provide limited information about cardiac mechanical function. In contrast, the quantitative impact of additional biosignals, such as ballistocardiography (BCG), which reflects the body's minute mechanical responses to cardiac ejection, remains unclear. In this study, we generated synthetic PPG, APW, and BCG signals from a unified whole-body cardiovascular circulation model and evaluated the complementary contribution of BCG to probabilistic cardiovascular biomarker estimation. We estimated posterior distributions of cardiovascular biomarkers using neural posterior estimation and simulation-based inference. The results showed that adding BCG signals significantly improved estimation performance. Furthermore, even in ill-posed cases where PPG and APW alone produced multimodal posterior distributions, adding BCG yielded unimodal posterior distributions. These findings provide fundamental insights into signal selection for estimating cardiovascular dynamics.
\end{abstract}

\begin{IEEEkeywords}
PPG signal, BCG signal, non-invasive biosignals, simulation-based inference, cardiovascular modeling, neural posterior estimation
\end{IEEEkeywords}

\section{Introduction}
\label{sec: introduction}
Recently, the number of patients with cardiovascular diseases, such as heart failure, has increased due to an aging global population, and this trend is expected to continue for decades \cite{Kazi2024-ad}. Furthermore, the medical costs associated with heart failure treatment are estimated to increase four- to five-fold over the next 30 years\cite{Kazi2024-ad}, representing a significant social burden. Once heart failure progresses to a certain extent, it becomes irreversible and tends to recur, which significantly worsens patients' quality of life\cite{Metra2023-yn}.

\textcolor{black}{
For the prevention and ultra-early detection of heart failure, routine and continuous monitoring of physiological biomarkers such as cardiac output, vascular resistance, and ejection fraction is effective. In fact, these indicators play a crucial role in interpreting the pathophysiology of progressive heart failure, assessing disease severity, determining treatment strategies, and monitoring treatment efficacy. Significant advances in measurement technology have made it possible to measure these indicators with high precision; however, in general, many of these require invasive procedures or specialized measurements. Consequently, it is not practical to continuously perform these high-cost measurements during the prevention phase of heart failure. Therefore, in recent years, technologies that predict physiological biomarkers from non-invasive biosignals---which can be measured using wearable devices or cuffless sensors---have been attracting attention \cite{Allen2007-tg, Kim2016-pc}. For example, non-invasive biosignals such as photoplethysmography (PPG), arterial pulse wave (APW), and ballistic electrocardiography (BCG) are being recognized as important clues for the early detection of heart failure \cite{Myhre2024-ec}. If it becomes possible to estimate clinically interpretable cardiovascular biomarkers from these non-invasive biosignals, this could contribute to the ultra-early detection and prevention of heart failure \cite{Behrmann2025-ji, Manduchi2024-yw, Wehenkel2023-mi}.
}

For the inverse estimation of cardiovascular biomarkers from noninvasive biosignals, several approaches combining cardiovascular simulation and machine learning have been proposed \cite{Bonnemain2021-jj, Jin2021-qf}.  Although these studies demonstrated the feasibility of estimating cardiovascular biomarkers from biological signals, they mainly treated the inverse mapping as a deterministic regression problem. However, inverse problems in the cardiovascular system are inherently non-deterministic because different combinations of physiological parameters can produce similar observable biosignals \cite{Wehenkel2023-mi}. Therefore, a single point estimate is often insufficient to characterize the multiple plausible solutions and the associated uncertainty. 

Simulation-based inference (SBI) provides a probabilistic framework for addressing inverse problems by using simulator-generated data to infer the posterior distribution of target biomarkers conditioned on observed signals \cite{Cranmer2020-lm}. In contrast to deterministic regression approaches, SBI aims to estimate not only a representative value but also the uncertainty and possible variability of the parameters that are consistent with the observations. This property makes SBI particularly suitable for medical applications, where obtaining large-scale clinical data is difficult, and for complex physiological systems, where analytical inverse estimation is challenging \cite{Behrmann2025-ji}. In this context, Wehenkel et al. (2023) proposed an SBI-based method using neural posterior estimation (NPE) to estimate posterior probability distributions of cardiovascular biomarkers from PPG and arterial pressure signals \cite{Wehenkel2023-mi}. 

Wehenkel et al. performed the inverse estimation using synthetic PPG and APW signals generated from data created by Carlson et al.\cite{Charlton2019-py}. Although PPG and APW reflect different physiological quantities, these signals exhibit strong similarities in both the time and frequency domains\cite{Martinez2018-ap, Ibtehaz2022-ex}. This strong correlation suggests that the information provided by PPG and APW for estimating cardiovascular biomarkers may be partially redundant. In particular, because PPG and APW mainly reflect peripheral vascular volume and pressure waveforms, they may provide limited information about central cardiac ejection dynamics and cardiac mechanical function \cite{Mukkamala2010-bo}. Although BCG has been used to estimate cardiovascular indices such as heart rate, blood pressure, and cardiac output, most existing studies have focused on specific indices, using empirical or regression-based approaches \cite{Tramontano2023-tm, Gonzalez-Landaeta2022-li}. The complementary contribution of BCG to the probabilistic estimation of multiple cardiovascular biomarkers, particularly in combination with PPG and APW, remains insufficiently investigated. Therefore, this study evaluates whether incorporating BCG into an SBI framework improves the accuracy, uncertainty characterization, and calibration of cardiovascular biomarker estimation.

In this study, we investigated whether BCG provides complementary information for the probabilistic estimation of the posterior distributions of cardiovascular biomarkers when combined with conventional PPG and APW signals. A key requirement for this evaluation is a unified physiological model that can generate PPG, APW, and BCG from the same underlying cardiovascular state and parameter set, while covering both cardiac function and vascular dynamics. To this end, we develop a closed-loop 0D-1D cardiovascular model by representing the heart as a zero-dimensional compartment model and the vascular system as a zero-dimensional model. This integrated model enables simultaneous generation of PPG, APW, and BCG signals, along with ground-truth biomarkers for both cardiac and vascular function. Using the simulated dataset, we train neural posterior estimation within the SBI framework and quantitatively evaluate how different signal combinations, particularly the inclusion of BCG, affect the accuracy, uncertainty characterization, and calibration of cardiovascular biomarker estimation.

\section{Methods}
The cardiovascular closed-loop model used in this study follows Liang et al. \cite{Liang2009-km}, combining a 1D vascular model with 0D cardiac and peripheral models. This multi-scale approach enables efficient simulation of pulse wave propagation in the large arteries while capturing global hemodynamic behavior through the lumped-parameter models. A schematic illustration of the model is shown in Fig. 2. The details of each component are described in the following subsections.
\begin{figure}[thbp]
\centering
\includegraphics[scale=0.23]{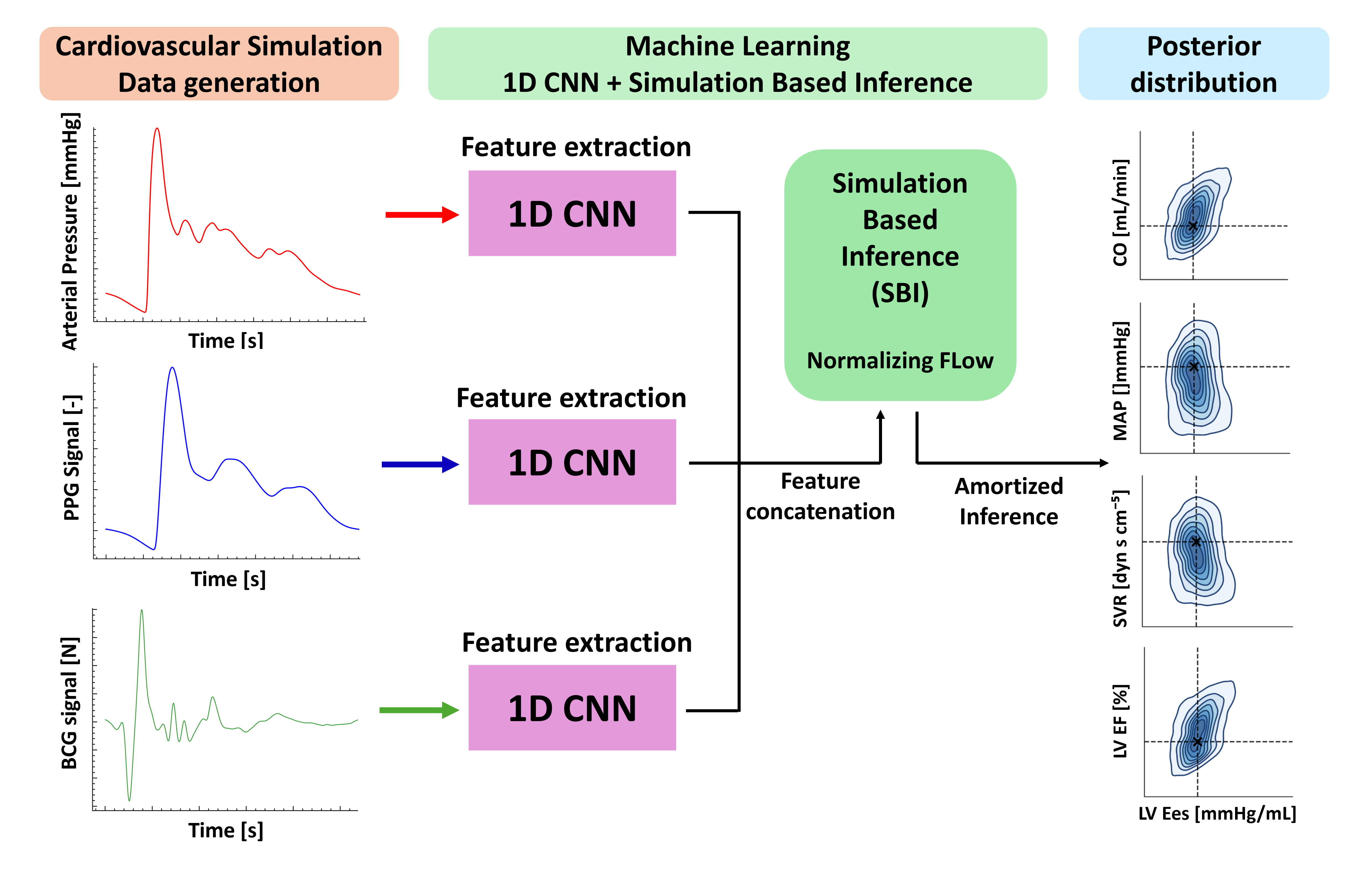}
\vspace{-8mm}
\caption{Framework for estimating cardiovascular biomarkers from non-invasive biosignals. Left: A 0D-1D cardiovascular simulator generates arterial pressure wave (APW), PPG, and BCG signals. Middle: Multiple signal combinations are constructed from the generated signals, and each combination is fed into one-dimensional convolutional neural networks (1D CNNs) to extract features. The extracted features are integrated and used as inputs to a normalizing-flow-based posterior estimator. Right: Posterior distributions of cardiovascular biomarkers estimated via simulation-based inference (SBI).}
\vspace{-3mm}
\label{fig:fig1}
\end{figure}

\subsection{1D Vascular Model}
In this study, the human arterial system was modeled using 55 vessels based on a previous study \cite{Liang2009-km, Rabineau2021-ja}. 
\textcolor{black}{
We will first describe the models within each vessel and then extend them into an integrated model of all 55 vessels.
The core state parameters of each vessel 
are the volumetric flow rate $Q(x, t)$, blood pressure $P(x, t)$, and vascular cross-sectional area $A(x, t)$ at axial position $x$ and time $t$ in each artery. 
}
The governing equations for blood flow are the cross-sectional averaged continuity equation and the Navier-Stokes equation: 
\begin{equation}
    \frac{\partial A}{\partial t}+\frac{\partial Q}{\partial x} = 0,
\end{equation}
\begin{equation}
    \frac{\partial Q}{\partial t}+\frac{\partial}{\partial x}\left(\frac{Q^2}{A}\right)+\frac{A}{\rho}\frac{\partial P}{\partial x} = -\frac{2\pi \nu r}{\delta}\frac{Q}{A},
\end{equation}
where 
$\rho$ is the blood density, $\nu$ is the kinematic viscosity, $r$ is the vessel radius, and $\delta$ is the thickness of the boundary layer \cite{Olufsen2000-sa}. 
These parameters $\rho$, $\nu$, and $r$ can be considered constants that reflect the physiological characteristics of the individual. The settings used in our model-based simulations are described in detail in Section \ref{sec:comp_setup}. 

The additional relationship between pressure and the cross-sectional area is expressed by the state equation, given by

\begin{equation}
P(x,t)-P_0=\frac{4}{3}\frac{Eh}{r_0}\left(1-\sqrt{\frac{A_0}{A}}\right),
\end{equation}
\textcolor{black}{where $P_{0}$ is the reference pressure at which the geometry is described by the cross-sectional area $A_{0}$, and ${Eh}/{r_0}$ is the elasticity parameter which will be detailed below.}
Specifically, 
${Eh}/{r_0}$ was expressed by the empirical equation \cite{Olufsen2000-sa} as

\begin{equation}
\frac{Eh}{r_0}=k_1 \exp\left(k_2 r_0\right)+k_3,
\end{equation}
\textcolor{black}{where the radius $r_{0}$ along the length of each artery was given as an exponentially tapered distribution using the inlet radius $R_u$, outlet radius $R_d$, and vessel length $L$ as follows:  
\begin{equation} 
r_0(x) = R_u \exp \left(\log\left(\frac{R_d}{R_u}\right)\frac{x}{L}\right).
\end{equation}
} 
\noindent 
The governing equations were discretized using the two-step Lax-Wendroff scheme \cite{Liang2009-km, Liu2024-ii, Olufsen2000-sa, Kolachalama2007-vf}. 

\textcolor{black}{
Now that we have constructed models for each vessel, we will extend them into a unified model of all 55 vessels.
We introduce the index set $\textsf{vasc}:=\{1,2,\dots,55\}$ to represent the compartments of the vessels, and denote volumetric flow rate and blood pressure of the $j$-th $(j\in\textsf{vasc})$ vessel by $Q_{i}(x, t)$ and $P_{i}(x, t)$, respectively.
The laws of mass conservation and pressure continuity must hold at the arterial bifurcation:
\begin{equation}
Q_{i}(x_{i},t) = Q_{\textrm{down1}(i)}(x_{i},t)+Q_{\textrm{down2}(i)}(x_{i},t),
\end{equation}
\begin{equation}
P_{i}(x_{i},t) = P_{\textrm{down1}(i)}(x_{i},t) = P_{\textrm{down2}(i)}(x_{i},t),
\end{equation}
where $x_{i}$ represents the terminal position of the $i$-th vessel, while $\textrm{down1}(i)$ and $\textrm{down2}(i)$ denote the indices of the two vessels bifurcating from the $i$-th vessel.
}
\textcolor{black}{In practice, these conditions}
were satisfied iteratively using Newton's method based on the ghost point method \cite{Liu2024-ii, Liang2009-km} at each time step.

Regarding the boundary conditions, the inlet boundary condition was imposed by coupling with a 0D cardiac model described in Section~\ref{sec:0D Cardiac Model} using extrapolating Riemann invariants\cite{Liang2009-km}, and the outflow boundary of each terminal vessel was coupled to the 0D peripheral vascular model through a three-element \textcolor{black}{Windkessel} model\cite{Liang2009-km, Kolachalama2007-vf}. 

\subsection{0D Cardiac and Peripheral Model}
\label{sec:0D Cardiac Model} 
The cardiac model consists of six compartments---
\textcolor{black}{
Left ventricle (lv); 
Right ventricle (rv);
Left atrium (la);
Right atrium (ra);
Pulmonary artery (pa);
Pulmonary vein (pv)
}
---and is modeled as a closed-loop circuit including the pulmonary circulation \cite{Liang2009-km}. 
\textcolor{black}{
The core state parameters of the $j$-th $(j\in\ \textsf{card}:=\{\text{lv}, \text{rv}, \text{la}, \text{ra}, \text{pa}, \text{pv}\})$ compartment are the volume $V_{j}(t)$ 
and pressure $P_{j}(t)$.
Here
$Q_{\text{mt}}$, 
$Q_{\text{av}}$, 
$Q_{\text{tc}}$, and
$Q_{\text{pv}}$
represent the flow rates through the mitral valve (mt), aortic valve (av), tricuspid valve (tc), and pulmonary valve (pv), respectively.
The model consists of 10 variables $U:=\{Q_{\text{mt}}, Q_{\text{av}}, Q_{\text{tc}}, Q_{\text{pv}}, V_{\text{lv}}, V_{\text{rv}}, V_{\text{la}}, V_{\text{ra}}, V_{\text{pa}}, V_{\text{pv}}\}$.
In a high-level formal representation, it is described as the following system of the ordinary differential equation:
\begin{equation}
\frac{dU}{dt} = F(U; e(t), \mathcal{C}),
\end{equation}
where $F$ is the set of system equations, $e(t):=\{e_{a}(t), e_{v}(t)\}$ is the set of cardiac driver functions, and $\mathcal{C}$ is the set of constant parameters.
For readability, we denote constants contained in $\mathcal{C}$ using a superscript bar, as in $\bar{A}$, for example.
The system equations consist of \emph{(a) volume change} and \emph{(b) flow rate}, \emph{(c) septum free wall volume}, and \emph{(d) peripheral circulation}. 
An overview of the model is provided, for example, in Appendix A of \cite{Grigorian2024-ng}. Due to space constraints, we will focus here on the most important modules.
}

\textcolor{black}{
\emph{(a) Volume change} - 
The system equations $F$ include the relationship between volume changes and in-out flows for each chamber, as illustrated below:
\begin{equation}\label{volume_change}
\frac{dV_{\text{lv}}}{dt}=Q_{\text{mt}}-Q_{\text{av}}, \ \
\frac{dV_{\text{rv}}}{dt}=Q_{\text{tc}}-Q_{\text{pv}}, \ \ \dots
\end{equation}
}
\textcolor{black}{
\emph{(b) Flow rate} - 
The system equations $F$ also include the governing equations for flow rates.
}
The flow rate through the $j$-th $(j\in\{\text{mt},\text{av},\text{tc},\text{pv}\})$ valve is governed by inductance $\bar{L}_j$, resistance $\bar{R}_j$, and a flow-dependent resistance coefficient $\bar{B}_j$: 
\textcolor{black}{
\begin{equation}
\bar{L}_j\frac{dQ_j}{dt} = P_{\text{up}(j)} - P_j - \bar{R}_j Q_j - \bar{B}_j |Q_j| Q_j,
\end{equation}
}
where \textcolor{black}{$\text{up}(j)$ the index of the compartment immediately upstream of the $j$-th compartment when blood is flowing in the physiological direction.} 
Backflow was prevented by setting $Q_j=0$ whenever $Q_j<0$.

\textcolor{black}{
\emph{(c) Septum free wall volume} - 
To simulate the pressures and interactions between the left and right ventricles more accurately, the notions of \emph{septum free wall} (spt), \emph{left ventricular free wall} (lvf), and \emph{right ventricular free wall} (rvf) are commonly used \cite{Liang2009-km,Grigorian2024-ng}. For each volume $V_{\cdot\in\{\text{spt},\text{lvf},\text{rvf}\}}(t)$ and pressure $P_{\cdot\in\{\text{spt},\text{lvf},\text{rvf}\}}(t)$, the following conditions are imposed.
First, by separating the volumes of the ventricular septum and the free walls of the left and right ventricles, this approach avoids the discrepancies caused by the ventricular septum, which is shared by both ventricles. Specifically, $V_{\text{lvf}}=V_{\text{lv}}-V_{\text{spt}}$ and $V_{\text{rvf}}=V_{\text{rv}}-V_{\text{spt}}$. 
Next, since the ventricles are located within the pericardium, a new term, pericardial pressure $P_{\text{peri}}(t)$, is introduced, and the following relationship holds:
$P_{\text{lv}}=P_{\text{lvf}}+P_{\text{peri}}$ and $P_{\text{rv}}=P_{\text{rvf}}+P_{\text{peri}}$.
}
\textcolor{black}{Then,} the pressure-volume relationship in the ventricles and atria is described by a time-varying elastance model \cite{Liang2009-km}. Taking the left ventricular free wall as an example, the pressure is given by \cite{Grigorian2024-ng} \textcolor{black}{as follows: 
\begin{equation}
\begin{split}
P_{\text{lvf}} &= e_{a}(t) \bar{E}_{\text{es,lvf}}
\left(V_{\text{lvf}} - \bar{V}_{\text{d,lvf}}\right) \\
&+ \left(1 - e_{a}(t)\right) \bar{P}_{\text{0,lvf}}
\left(e^{\bar{\lambda}_{\text{lvf}}(V_{\text{lvf}} - \bar{V}_{\text{0,lvf}})} - 1\right),
\end{split}
\end{equation}}
\textcolor{black}{
\noindent where $\bar{E}_{\text{es,lvf}}$ is left ventricular end systolic elastance, $\bar{V}_{\text{d,lvf}}$ is unstressed left ventricular volume, $\bar{P}_{\text{0,lvf}}$ is left ventricular EDPVR gradient, $\bar{V}_{\text{0,lvf}}$ is zero-pressure left ventricular volume,  $\bar{\lambda}_{\text{lvf}}$ is left ventricular EDPVR curvature, and the cardiac driver function $e_{a}(t)$ is given in Liang et al. \cite{Liang2009-km}. 
Similar pressure-volume formulations were applied to the other chambers.
}
\textcolor{black}{Finally, the}
pressure equilibrium condition among the left and right free walls and the septum is given by
$
\textcolor{black}{
P_{\text{spt}} - P_{\text{lvf}} + P_{\text{rvf}} = 0.
}
$
In practice, these conditions are solved using Newton's method at each time step \cite{Grigorian2024-ng}.

\emph{(d) Peripheral circulation} - 
Pericardial constraint is expressed as a function of the total pericardial volume $V_{\text{pcd}}(t) := V_{\text{lv}}(t)+V_{\text{rv}}(t)+V_{\text{la}}(t)+V_{\text{ra}}(t)$, given by
\textcolor{black}{
\begin{equation}
P_{\text{peri}} = \bar{P}_{\text{0,pcd}}\!\left(e^{\bar{\lambda}_{\text{pcd}}(V_{\text{pcd}}-\bar{V}_{\text{0,pcd}})}-1\right)
           + \bar{P}_{\text{th}},
\end{equation}
where $\bar{P}_{\text{0,pcd}}$ is the zero-volume pericardium pressure,  $\bar{\lambda}_{\text{pcd}}$ is 
$V_{\text{0,pcd}}$ is the zero-pressure pericardium volume, and $\bar{P}_{\text{th}}$ is the thoracic cavity pressure. 
}
The peripheral systemic circulation is represented by four compartments---capillary, venule, vein, and vena cava---forming a closed loop model in which blood flows from the Windkessel models connected to each arterial terminal into the venous compartments and returns to the right atrium via the vena cava. 

\begin{figure}[thbp]
\centering
\includegraphics[scale=0.31]{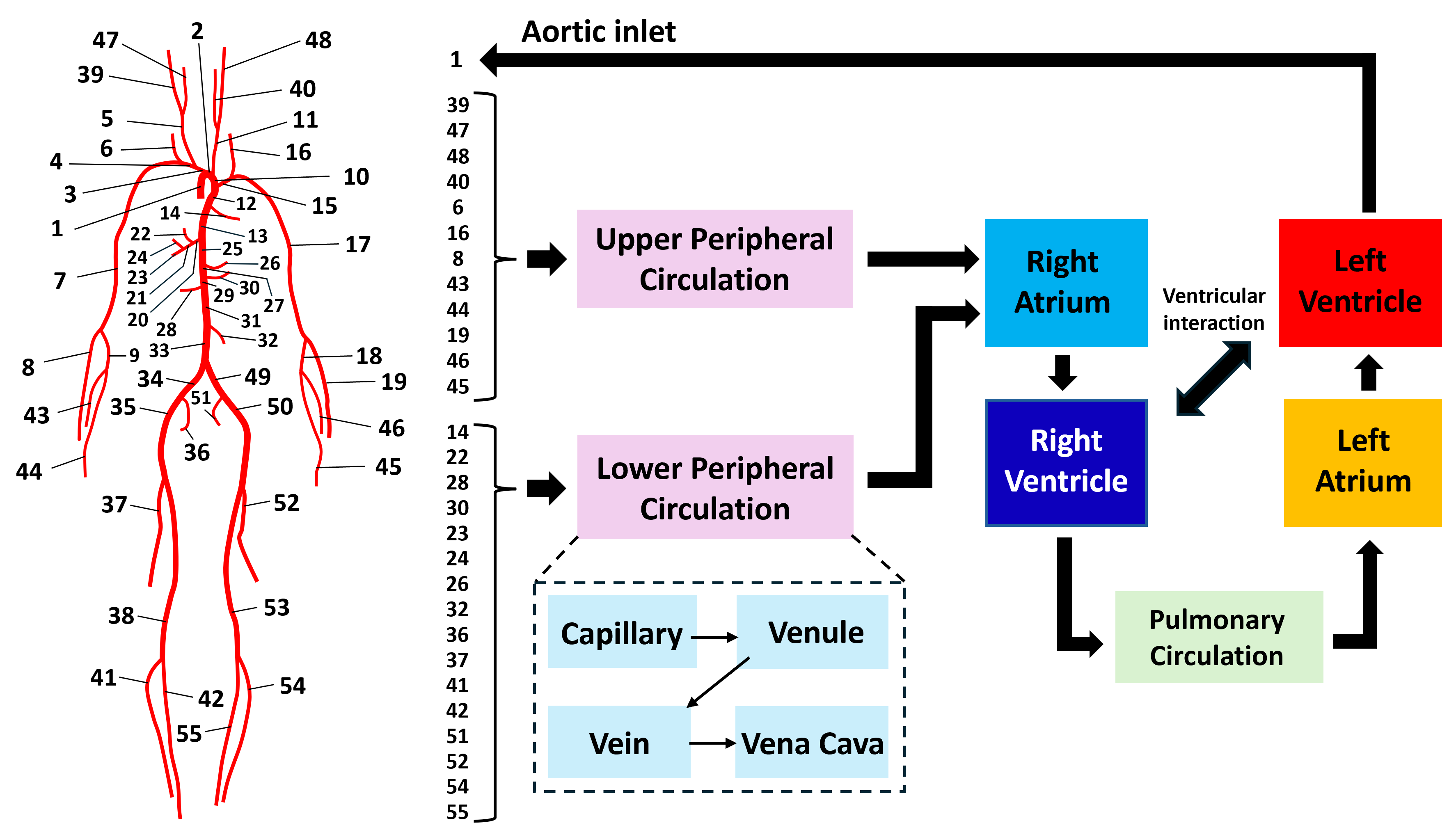}
\vspace{-3mm}
\caption{Schematic illustration of the cardiovascular closed-loop model used in this study. The vascular tree comprises 55 arteries, modeled using the 1D formulation, while the peripheral and cardiac circulations are modeled using the 0D formulation. The 1D and 0D models are coupled at each arterial terminal via a three-element Windkessel model. The model structure is based on Liang et al. \cite{Liang2009-km}. }
\label{fig:fig2}
\end{figure}

\subsection{Computational Setup}\label{sec:comp_setup}
For the 1D vascular model, the time step and mesh size were set to
$\Delta t = 1\times 10^{-5}$~s, $\Delta x =0.1$~cm, respectively to satisfy the
Courant-Friedrichs-Lewy (CFL) condition. The blood density and the kinematic viscosity were set to $\rho = 1.06\, \mathrm{g \cdot cm^{-3}}$ and $\nu = 4.43 \, \mathrm{cm^{2}\, s^{-1}}$, respectively.
The vessel lengths were taken from Rabineau et al. \cite{Rabineau2021-ja}, and the inlet and outlet radii of each
segment and the Windkessel parameters ($R_1$, $R_2$, and $C_T$)
were taken from Liang et al. \cite{Liang2009-km}.
For the 0D cardiac and peripheral model, the same time step was
used, and all governing equations were integrated using the
fourth-order Runge--Kutta method.
Baseline cardiac and peripheral parameters were taken from
Liang et al. and Grigorian et al. \cite{Liang2009-km, Grigorian2024-ng}.
The simulation was run for 30 cardiac cycles to reach a periodic
steady state, and the last five cycles were used for training.
All computations were performed on an Intel Xeon w5-2455X.
 
\subsection{Signal Generation}
We generated APW, PPG, and BCG signals from the cardiovascular closed-loop model described above.
The APW signal was taken as the pressure waveform at the midpoint of the radial artery.
The PPG signal was obtained as the blood volume change within the Windkessel model connected to the terminal of the ulnar artery, computed as \cite{Charlton2019-py}
\begin{equation}
\text{PPG}(t)=\int_0^t Q_{1D}(t') - Q_{out}(t') dt',
\end{equation}
where $Q_{1D}$ is the flow rate entering the Windkessel model from the 1D vessel, and $Q_{out}$ is the flow rate leaving the Windkessel model.  

The BCG signal was computed following Rabineau et al. \cite{Rabineau2021-ja}, using the three-dimensional positions of the four cardiac chambers and the vascular tree, along with the flow rate in each compartment. The velocity-BCG signal is given by 
\begin{equation}
\mathbf{BCG}_{\mathrm{vel}}(t)
= -\frac{\rho}{W_b}\sum_{j\in \textsf{vasc}\cup\textsf{card}} \hat{Q}_j(t)
  \left(\mathbf{G}_j - \mathbf{G}_{\text{up}(j)}\right)
\end{equation}
where $\hat{Q}_{j}(t):=Q_{j}(L/2,t)$ (for $j\in \textsf{vasc}$); $\hat{Q}_{j}(t):=Q_{j}(t)$ (for $j\in \textsf{card}$), and 
$W_b$ is the body weight, 
\textcolor{black}{
and $\mathbf{G}_j$ is the position vector of the center of mass 
of the $j$-th compartment.
}

\begin{table}[htbp]
\vspace{-3mm}
\centering
\scriptsize
\caption{Model parameters and ranges used in the simulations.}
\label{table1}
\begin{tabular}{lll}
\hline
\textbf{Parameter} & \textbf{Description} & \textbf{Range} \\ \hline
cardiac cycles & Number of cardiac cycles & $[40,150]$ \\
$k_1$ & Coefficient in the Eq.(6) & $[1.5\times 10^6,5\times10^6]$ \\ 
$k_2$ & Coefficient in the Eq.(6) & $[-15.0,-5.0]$ \\ 
$k_3$ & Coefficient in the Eq.(6) & $[1.5\times 10^5,5\times 10^5]$ \\
$E_{\mathrm{es,lvf}}$ & Left ventricular end-systolic elastance & $[0.5,5.0]$ \\ 
$R_2$ & Scaling factor for Windkessel $R_2$ & $[0.9, 2.0]$ \\
$C_t$ & Scaling factor for Windkessel $C_t$ & $[0.5, 2.0]$ \\
\hline
\end{tabular}
\vspace{-2mm}
\end{table}

We generated simulator-based training datasets that include patients with heart failure and healthy individuals by varying the LV Ees value, with the aim of preventing and detecting heart failure.
Specifically, we generated 10,000 sets of input parameters and a dataset comprising three signal types (PPG, APW, and BCG) by randomly sampling the 7 parameters shown in Table \ref{table1} within their respective ranges. Finally, these signals were fed into a convolutional neural network (CNN) to extract features. The CNN architecture follows a previous study \cite{Manduchi2024-yw}.

\subsection{Simulation-Based Inference (SBI)}
Simulation-based inference (SBI), also referred to as likelihood-free inference, is a framework for estimating posterior distributions when the likelihood function is intractable but forward simulation is available. Given observed biosignals $x$, SBI aims to estimate the posterior distribution of target cardiovascular quantities, $p(\phi|x)$. In contrast to conventional point-estimation approaches, such as surrogate models, SBI enables quantitative uncertainty estimation and can represent multimodal posterior distributions even for ill-posed inverse problems. 

In this study, following the previous study by Manduchi et al. \cite{Manduchi2024-yw}, SBI was applied to estimate the target cardiovascular physiological biomarkers from observed biosignals $x$. Here, $x$ denotes the observed dataset consisting of one or more signals selected from APW, PPG, and BCG. Let $\vartheta \in \Theta$ denote the full set of variables governing the simulator, and let $\phi = g(\vartheta)$ denote the target biomarkers of interest. Here, $\vartheta$ denotes vascular parameters, including radii, lengths, elasticity, and related parameters, as well as cardiac parameters, such as cardiac elastance and related parameters. This formulation enables the consistent treatment of not only direct input parameters, but also physiological indicators derived from simulation results. In this study, the target biomarkers $\phi$ were defined to include left ventricular end-systolic elastance (LV Ees), cardiac output (CO), heart rate (HR), mean arterial pressure (MAP), central venous pressure (CVP), mean pulmonary arterial pressure (mPAP), pulmonary capillary wedge pressure (PCWP), systemic vascular resistance (SVR), pulmonary vascular resistance (PVR), left and right ventricular ejection fractions (LV EF and RV EF), and left ventricular end-diastolic and end-systolic volumes (LV EDV and LV ESV).

Neural posterior estimation (NPE) was used as the SBI algorithm to estimate $p(\phi|x)$. Given the dataset $\mathscr{D}=\{(\phi_i, x_i)\}_{i=1}^M$, $M$ represents the number of samples, and we approximate the posterior distribution of the biomarkers $\phi$ given the observed signals $x$ by training a conditional density estimator $p_{\omega}(\phi|x)$, where $\omega$ denotes the trainable parameters of the estimator. In the framework of NPE, the density estimator was trained by solving the optimization problem as follows: 
\begin{equation}
\begin{aligned}
\omega^\star
&\in \arg \min_{\omega \in \Omega}
\mathbb{E}_{x}
\left[
\mathrm{\mathbb{KL}}\!\left(
p(\phi | x)\,\|\,p_{\omega}(\phi | x)
\right)
\right] \\
&\iff
\omega^\star
\in \arg \max_{\omega \in \Omega}
\mathbb{E}_{(\phi,x)}
\left[
\log p_{\omega}(\phi | x)
\right].
\end{aligned}
\end{equation}
That is, minimizing the difference between the true posterior distribution $p(\phi|x)$ and the approximate posterior distribution $p_{\omega}(\phi| x)$ in terms of the Kullback--Leibler divergence is equivalent to maximizing the expected log-likelihood. In practice, this optimization problem is implemented as the minimization of the negative log-likelihood as

\begin{equation}
\mathcal{L}_{\mathrm{NPE}}(\mathscr{D},\omega)
=
-\frac{1}{K}
\sum_{k=1}^{K}
\log p_{\omega}(\phi_i | x_i).
\end{equation}
Following the previous study, the conditional density estimator $p_{\omega}(\phi|x)$ was parametrized using a normalizing flow (NF) proposed by Rezende et al. \cite{Rezende2015-fi}. In NF, an invertible transformation with respect to $\phi$ is introduced by
\begin{equation}
z = f_{\omega}(\phi; x)
\end{equation}
which maps $\phi$ to a simple base distribution $p_z(z)$. In this study, an isotropic Gaussian distribution was used as a base distribution $p_z$. Then, according to the formula for change-of-variables, the conditional density and the loss of training using NF are given by \cite{Tabak2012-fb}
\begin{equation}
\log p_{\omega}(\phi|x)
=
\log p_z\!\left(f_{\omega}(\phi; x)\right)
+
\log \left|
J_{f_{\omega}}\left(\phi;x \right)
\right|,
\end{equation}
\begin{equation}
\begin{aligned}
\mathcal{L}&_{\mathrm{NF}}(\mathscr{D},\omega)
=\\
&-\frac{1}{K}
\sum_{k=1}^{K}
\left[
\log p_z\!\left(f_{\omega}(\phi_i; x_i)\right)
+\log \left| J_{f_{\omega}} \left(\phi_i ; x_i \right) \right|
\right].
\end{aligned}
\end{equation}
Here $f_{\omega}: \mathbb{R}^{k} \times \mathbb{R}^{D} \rightarrow \mathbb{R}^{k}$ is an invertible mapping that transforms the target quantities $\phi$ into the latent space conditioned on the observed signals $x$, and a neural network represents it. Moreover, the second term corresponds to the determinant of the Jacobian matrix and accounts for the local change in volume of the probability density under the transformation. The NF used in this study was implemented based on Papamakarios et al. \cite{NIPS2017_6c1da886}.

\subsection{Model Training Setup}
The SBI input is the extracted features from signals, obtained using a 1D CNN feature extractor (Fig. 1). The 1D CNN feature extractor consisted of five convolutional layers.  Each convolutional layer used a kernel size of 3 without padding and was followed by a ReLU activation function.  The first three convolutional layers used 40 output channels with a stride of 2. A max-pooling layer with a kernel size of 3 was then applied. Finally, two convolutional layers with 20 and 10 output channels, respectively, were used to obtain compact temporal features. 

For NPE, we used a fully connected neural network with three hidden layers, each containing 200 neurons. The model was trained using the Adam optimizer for up to 5000 epochs with a batch size of 128. Early stopping was applied based on the validation loss to prevent overfitting.  The initial learning rate was set to $1 \times 10^{-3}$. A ReduceLROnPlateau scheduler was used to adaptively reduce the learning rate based on the validation loss. Training was conducted on an NVIDIA RTX 4000 Ada GPU.

\begin{figure}[t!]
\vspace{-1mm}
\centering
\includegraphics[scale=0.72]{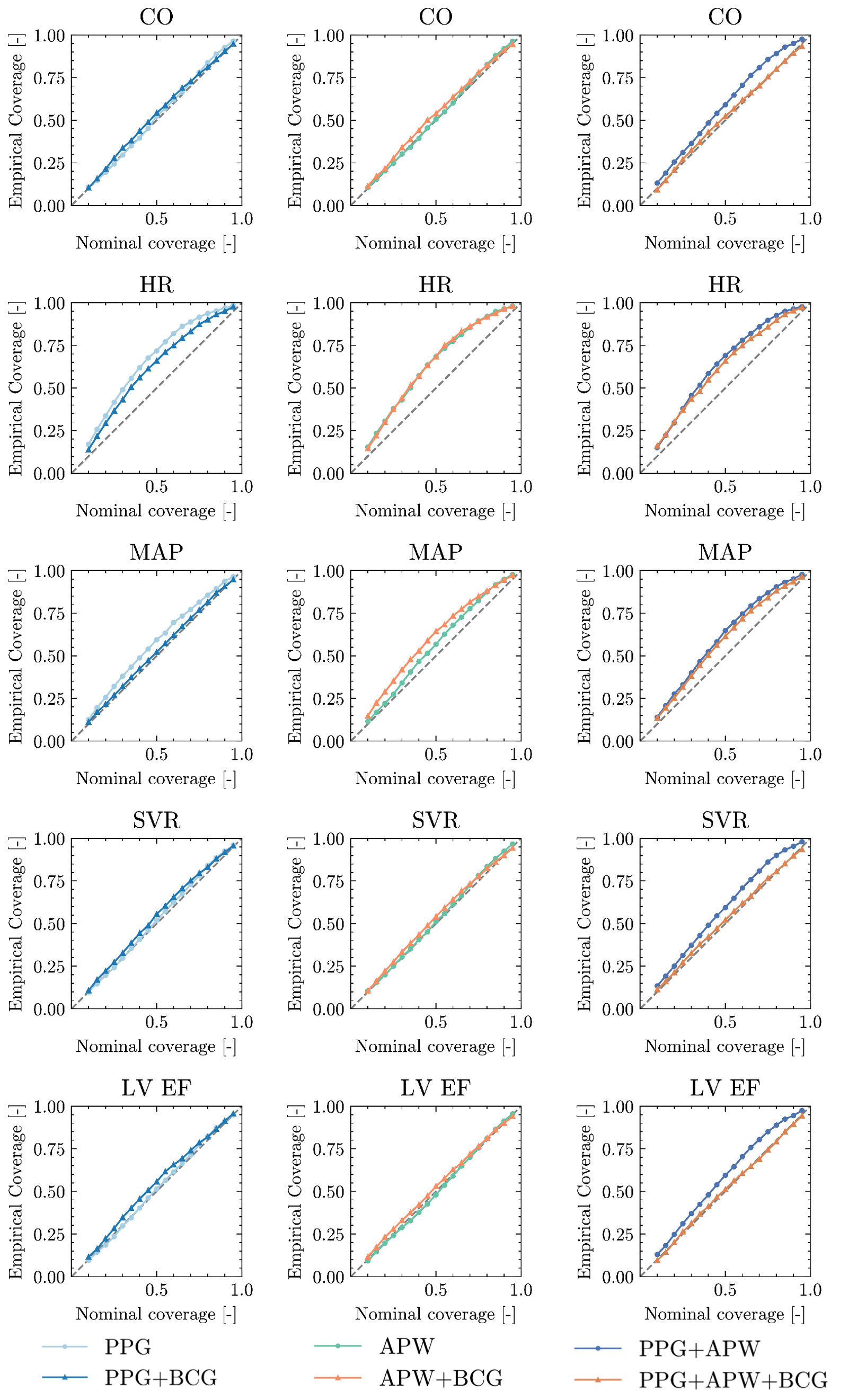}
\vspace{-5mm}
\caption{Coverage curves for representative cardiovascular biomarkers under each signal combination. The left, middle, and right columns compare PPG with PPG+BCG, APW with APW+BCG, and PPG+APW with PPG+APW+BCG, respectively. A well-calibrated posterior distribution should ideally follow the diagonal line.}
\label{fig:fig3}
\vspace{-3mm}
\end{figure}

\subsection{Model Evaluation}
To assess the calibration of the estimated posterior distributions, we employed the global coverage check, a widely used evaluation metric for SBI \cite{Deistler2025-lx}. This approach evaluates the validity of uncertainty quantification by examining whether true biomarker values fall within credible posterior intervals at nominal coverage levels. In addition, we quantified the calibration error using the mean absolute coverage error (MACE), defined as
\begin{equation}
\text{MACE}=\frac{1}{K}\sum_{k=1}^K |\hat{C}\left(c_k\right)-c_k|
\end{equation}
where $\hat{C}\left(c_k\right)$ denotes the empirical coverage at the nominal coverage level $c_k$, and $K$ is the number of coverage levels considered. In this experiment, K was set to 9. In addition to MACE, we evaluated the regression performance of the posterior point estimates. Specifically, for each test sample, the posterior sample with the highest density was taken as an approximate posterior mode estimate, and its agreement with the ground-truth biomarkers was assessed using the $R^2$ determinant.

\section{Results}
Fig. 3 shows the measured coverage for various signal combinations across representative cardiovascular biomarkers, including CO, HR, MAP, SVR, and LV EF. Additionally, Table \ref{table2} summarises the MACE scores for 13 biomarkers, including those related to pulmonary circulation. Compared to PPG alone, adding BCG consistently improved coverage. While the combination of PPG, APW, and BCG achieved low coverage error for most biomarkers, APW and BCG alone also demonstrated high coverage for some biomarkers. This indicates that increasing the number of input signals does not necessarily improve calibration.

\begin{table*}[htbp]
\vspace{-3mm}
\centering
\tiny
\caption{MACE scores for each cardiovascular parameter under different signal combinations. 
}
\label{table2}
  \begin{tabular}{llllllll}
    \hline
    \textbf{Cardiovascular Biomarkers} & \textbf{PPG} & \textbf{APW} & \textbf{BCG} & \textbf{APW+BCG} & \textbf{PPG+APW} & \textbf{PPG+BCG} & \textbf{PPG+APW+BCG} \\ \hline
    LV Ees & $0.0488\pm0.0349$ & \bm{$0.0375\pm0.0253$} & $0.0858\pm0.0174$ & $0.0814\pm0.0337$  & $0.0529\pm0.0369$ & $0.0632\pm0.0314$ & $0.0379\pm0.0222$\\
    CO     & $0.0518\pm0.0337$ & $0.0388\pm0.0257$ & $0.0364\pm0.0162$ & $0.0771\pm0.0346$  & $0.0543\pm0.0379$ & $0.0641\pm0.0322$ & \bm{$0.0306\pm0.0223$}\\ 
    HR     & $0.1440\pm0.0253$ & $0.1220\pm0.0159$ & $0.1140\pm0.0418$  & $0.1150\pm0.0147$   & $0.1200\pm0.0179$  & $0.1240\pm0.0175$  & \bm{$0.1090\pm0.0125$}\\ 
    MAP    & $0.0677\pm0.0311$ & $0.0679\pm0.0231$ & \bm{$0.0224\pm0.0123$} & $0.1030\pm0.0186$   & $0.0806\pm0.0276$ & $0.0493\pm0.0348$ & $0.0860\pm0.0117$\\
    CVP    & $0.0632\pm0.0372$ & $0.0556\pm0.0275$ & \bm{$0.0235\pm0.0133$} & $0.0304\pm0.0135$  & $0.0683\pm0.0344$ & $0.0539\pm0.0277$ & $0.0284\pm0.0148$\\				
    mPAP   & $0.0510\pm0.0337$ & $0.0389\pm0.0256$ & $0.0572\pm0.0157$ & $0.0667\pm0.0350$  & $0.0543\pm0.0383$ & $0.0648\pm0.0316$ & \bm{$0.0303\pm0.0209$}\\
    PCWP   & $0.0507\pm0.0340$ & $0.0384\pm0.0253$ & $0.0466\pm0.0166$ & $0.0647\pm0.0336$  & $0.0518\pm0.0381$ & $0.0610\pm0.0309$ & \bm{$0.0291\pm0.0204$}\\
    SVR    & $0.0541\pm0.0324$ & $0.0417\pm0.0252$ & \bm{$0.0201\pm0.0116$} & $0.0779\pm0.0342$  & $0.0544\pm0.0389$ & $0.0571\pm0.0209$ & $0.0393\pm0.0208$\\
    PVR    & $0.0718\pm0.0247$ & $0.0539\pm0.0194$ & $0.0406\pm0.0141$ & \bm{$0.0152\pm0.0083$} & $0.0570\pm0.0218$ & $0.0381\pm0.0287$ & $0.0175\pm0.0063$\\
    LV EF  & $0.0512\pm0.0341$ & $0.0390\pm0.0258$ & $0.0526\pm0.0131$ & $0.0725\pm0.0339$  & $0.0544\pm0.0372$ & $0.0693\pm0.0326$ & \bm{$0.0306\pm0.0208$}\\	
    RV EF  & $0.0485\pm0.0335$ & $0.0394\pm0.0262$ & $0.0568\pm0.0196$ & $0.0612\pm0.0336$  & $0.0533\pm0.0392$ & $0.0611\pm0.0322$ & \bm{$0.0333\pm0.0202$}\\
    LV EDV & $0.0515\pm0.0342$ & $0.0392\pm0.0258$ & $0.0600\pm0.0164$ & $0.0717\pm0.0325$	& $0.0540\pm0.0386$	& $0.0695\pm0.0352$ & \bm{$0.0298\pm0.0210$}\\
    LV ESV & $0.0510\pm0.0347$ & $0.0385\pm0.0260$ & $0.0531\pm0.0137$ & $0.0732\pm0.0336$  & $0.0535\pm0.0378$ & $0.0701\pm0.0341$ & \bm{$0.0299\pm0.0194$}\\
    \hline
  \end{tabular}
\end{table*}

\begin{table*}[htbp]
\vspace{-3mm}
\centering

\caption{$R^2$ values obtained from regression analysis between the posterior mode estimates and the ground-truth values for each cardiovascular parameter under different signal combinations. 
}
\label{table3}
\tiny
  \begin{tabular}{llllllll}
    \hline
    \textbf{Cardiovascular Biomarkers} & \textbf{PPG}   & \textbf{APW}         & \textbf{BCG}  & \textbf{APW+BCG}    & \textbf{PPG+APW} & \textbf{PPG+BCG}    & \textbf{PPG+APW+BCG} \\ \hline
    LV Ees  & $0.805±0.070$ & $0.846±0.017$ & $0.989±0.006$ & $0.993±0.002$ & $0.857±0.019$ & $0.985±0.005$      & \bm{$0.994±0.001$} \\
    CO      & $0.935±0.019$ & $0.939±0.006$ & $0.995±0.004$ & $0.998±0.000$ & $0.951±0.007$ & $0.996±0.001$      & \bm{$0.999±0.000$}\\
    HR      & $0.995±0.005$ & \bm{$1.000±0.000$} & \bm{$1.000±0.001$} & \bm{$1.000±0.000$} & \bm{$1.000±0.000$}  & $0.999±0.001$ & \bm{$1.000±0.000$}\\
    MAP     & $0.995±0.004$ & $0.998±0.000$ & $0.983±0.031$ & \bm{$1.000±0.000$} & $0.999±0.000$ & $0.998±0.001$ & \bm{$1.000±0.000$}\\
    CVP     & $0.987±0.006$ & $0.990±0.001$ & $0.986±0.020$ & $0.998±0.001$ & $0.993±0.001$ & $0.997±0.001$      & \bm{$0.999±0.000$}\\			
    mPAP    & $0.852±0.052$ & $0.876±0.013$ & $0.991±0.006$ & $0.996±0.002$ & $0.896±0.021$ & $0.990±0.005$      & \bm{$0.997±0.001$}\\
    PCWP    & $0.848±0.051$ & $0.871±0.014$ & $0.991±0.006$ & $0.996±0.002$ & $0.891±0.021$ & $0.990±0.004$      & \bm{$0.997±0.001$}\\
    SVR     & $0.741±0.068$ & $0.759±0.027$ & $0.842±0.262$ & $0.989±0.005$ & $0.799±0.030$ & $0.975±0.007$       & \bm{$0.991±0.002$}\\
    PVR     & $0.988±0.005$ & $0.992±0.001$ & $0.997±0.004$ & \bm{$0.999±0.000$} & $0.992±0.003$ & $0.998±0.001$  & \bm{$0.999±0.000$}\\
    LV EF   & $0.862±0.052$ & $0.887±0.013$ & $0.993±0.005$ & \bm{$0.997±0.001$} & $0.907±0.013$ & $0.993±0.003$ & \bm{$0.997±0.000$}\\
    RV EF   & $0.883±0.044$ & $0.907±0.009$ & $0.992±0.006$ & $0.996±0.002$ & $0.922±0.012$ & $0.991±0.005$      & \bm{$0.997±0.001$}\\
    LV EDV  & $0.878±0.042$ & $0.896±0.011$ & $0.995±0.003$ & \bm{$0.997±0.001$} & $0.914±0.012$ & $0.993±0.003$ & \bm{$0.997±0.001$}\\
    LV ESV  & $0.865±0.051$ & $0.889±0.012$ & $0.993±0.005$ & \bm{$0.997±0.001$} & $0.908±0.012$ & $0.993±0.003$ & \bm{$0.997±0.000$}\\
    \hline
  \end{tabular}
  \vspace{-3mm}
\end{table*}

The $R^2$ values obtained from the regression analysis between the posterior mode estimates and the ground-truth values for each cardiovascular biomarker are summarized in Table \ref{table3}. The combination of PPG, APW, and BCG achieved the best overall performance across all biomarkers.

Fig. 4 shows the box-plots of the credible interval widths for representative cardiovascular biomarkers under different signal combinations. Regarding the credible interval width, the combination of PPG, APW, and BCG exhibited the smallest widths across the five representative biomarkers. In contrast, the credible interval width obtained using two-signal combinations was not necessarily smaller than that obtained using a single signal input.

\begin{figure}[t!]
\centering
\includegraphics[scale=0.45]{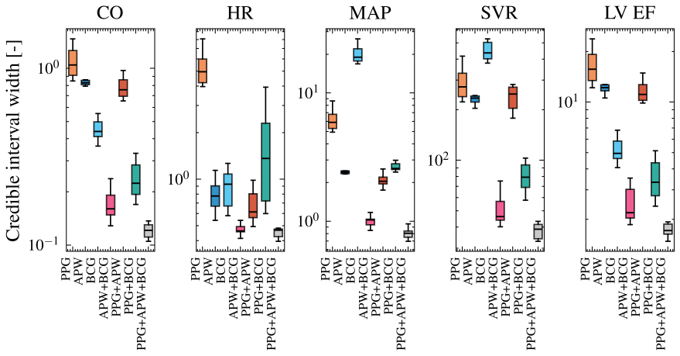}
\vspace{-3mm}
\caption{Box-plots of credible interval widths for representative cardiovascular biomarkers under different signal combinations: PPG, APW, BCG, APW+BCG, PPG+APW, PPG+BCG, and PPG+APW+BCG.}
\label{fig:fig4}
\vspace{-3mm}
\end{figure}

Fig. 5 shows the two-dimensional posterior distributions of five cardiovascular biomarkers for a representative sample, which showed multimodal distributions. Blue, orange, and green correspond to the posterior distributions obtained using PPG, PPG+APW, and PPG+APW+BCG, respectively. The diagonal elements show the one-dimensional marginal posterior distributions. Of all the signal combinations, the PPG+APW+BCG combination estimated posterior distributions that were most closely concentrated around the ground truth. Furthermore, incorporating the BCG signal transformed multimodal posterior distributions into unimodal ones.

\begin{figure}[t!]
\centering
\includegraphics[scale=0.47]{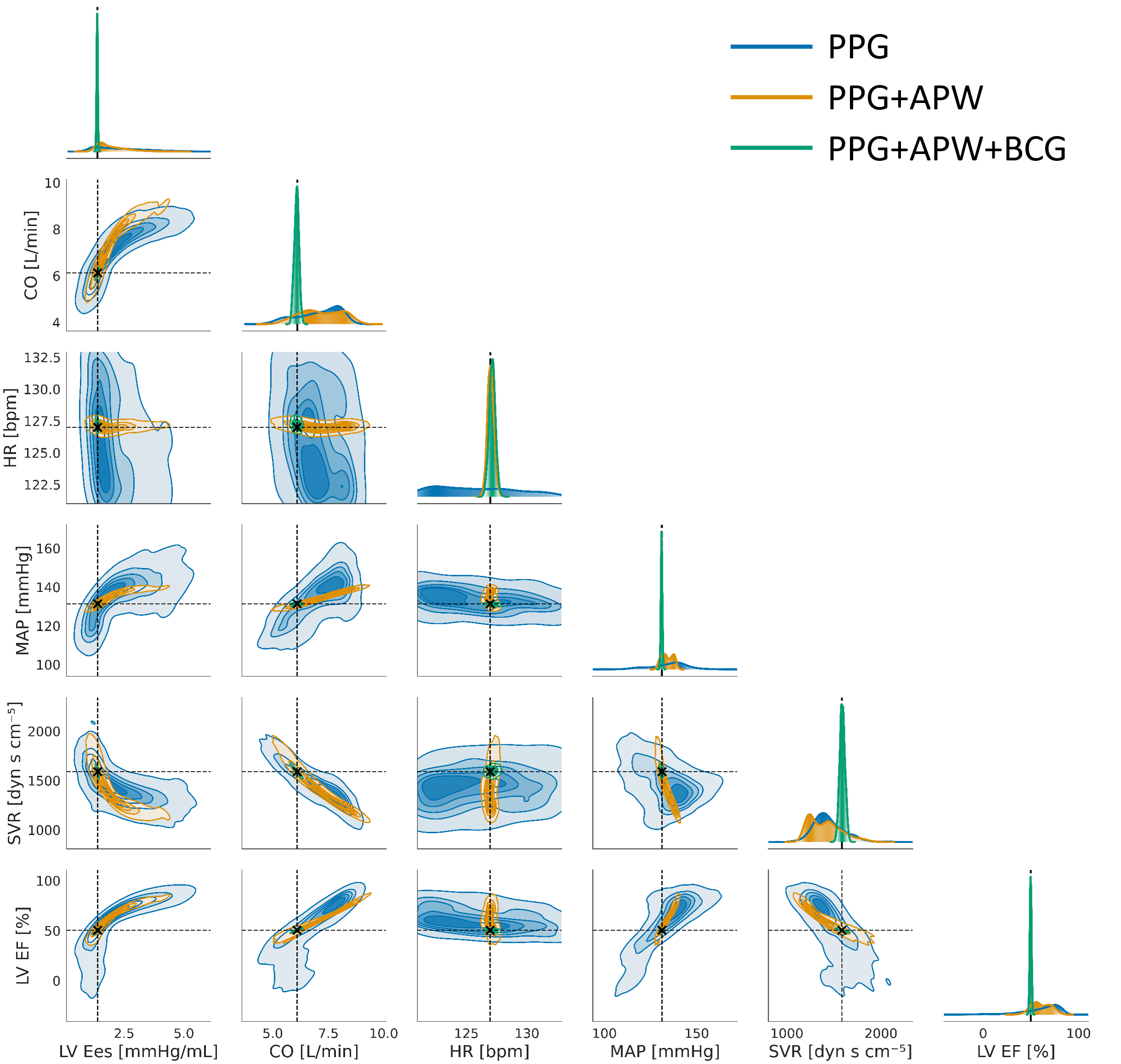}
\vspace{-5mm}
\caption{Posterior probability density distributions for a representative dataset. The diagonal panels show the one-dimensional marginal distributions, while the off-diagonal panels show the two-dimensional joint distributions. Blue, orange, and green denote PPG, PPG+APW, and PPG+APW+BCG, respectively.}
\label{fig:fig5}
\vspace{-3mm}
\end{figure}

\section{Discussion}
In this study, we quantitatively evaluated the impact of incorporating the BCG signal into the input modalities within a framework for cardiovascular biomarker estimation based on non-invasive biosignals. Using 10,000 synthetic datasets generated from a 0D-1D cardiovascular simulation, we demonstrated that improvements in inference performance across multiple cardiovascular biomarkers occurred when the BCG signal was incorporated. These results provide fundamental insights into the frameworks that integrate physical models with measurement data \cite{Sel2024-fy}.

On the other hand, Fig. 3 and Table \ref{table2} show that increasing the number of input signals does not necessarily improve calibration performance. The credible interval width of the posterior distribution decreases as the number of input signals increases (Fig. 4(A)). This result implies that, as the amount of information increases, the posterior distribution may become overly concentrated, leading to an underestimation of uncertainty.

Moreover, Fig. 5 shows that adding the BCG signal not only improves estimation accuracy but also affects the structure of uncertainty. In particular, under the PPG+APW condition, multimodal posterior distributions were observed in some samples, but they became unimodal when the BCG signal was incorporated. This result suggests that including additional observational signals improves differentiability in ambiguous estimation problems with multiple solutions.

The notable feature of this study is the use of an integrated 0D-1D model that accounts for interactions across the whole-body circulation, including the cardiac, pulmonary, and peripheral circulations. Conventional approaches have typically relied on simplified computational models, such as standalone 1D vascular models \cite{Nolte2022-ek} or 0D cardiovascular models \cite{Kuang2024-tt}. Moreover, previous studies \cite{Wehenkel2023-mi} that utilized existing datasets \cite{Charlton2019-py} have limitations in estimating cardiac function metrics and extending the underlying models. In contrast, our approach demonstrated the potential for a unified evaluation of cardiac function and pulmonary circulation dynamics by simultaneously estimating key cardiac biomarkers, such as LV Ees and LV EF, and pulmonary circulation parameters, including PVR and PCWP, from multiple non-invasive biosignals. Furthermore, because these indices reflect physiological conditions related to left and right ventricular preload and afterload\cite{Pinsky2016-xo}, the proposed method has the potential to provide a comprehensive evaluation of circulatory function based on cardiopulmonary interactions\cite{Alvarado2023-nd}. Additionally, this method does not require prior specification of the aortic inflow velocity, allowing it to naturally incorporate physiological feedback, such as hemodynamic changes associated with heart rate variability. This feature offers significant potential for future extension toward bio-digital twins \cite{Sel2024-fy}, particularly real-time bio-digital twins.

There are three main limitations in this study. First, this study relies solely on synthetic datasets generated by simulations. In general, misspecification may arise between real biological phenomena and simulation models \cite{Brynjarsdottir2014-ev}. Therefore, it is not guaranteed that the findings obtained in this study can be directly transferred to real clinical datasets. As a next step, it will be necessary to evaluate the applicability of this method in real-world settings by validating the parameter estimation performance using actual non-invasive measurement data. Second, the real-world datasets may include various noise distributions as well as temporally structured noise associated with the measurement environment. Therefore, it will be necessary to both evaluate the proposed method under more realistic noise models and develop methods that improve robustness to unknown noise distributions. To address these challenges, physics-guided inference methods, such as physics-informed neural networks \cite{Pilar2024-iz}, may be effective. Finally, the geometric structure used to generate BCG signals was based on the data from only one patient \cite{Rabineau2021-ja}. In practice, body size and vascular geometry can affect signal characteristics, and thus the model used in this study may not generalize to a broader population. To generate more realistic synthetic datasets, modeling based on the patient-specific geometries and physiological characteristics will be necessary. This will likely require access to real clinical datasets and collaboration with medical institutions to further extend the model. 

\section{Conclusion}
In this study, we evaluated the relationship between combinations of non-invasive biosignals and the inference of cardiovascular biomarkers. We demonstrated that integrating PPG, APW, and BCG signals improves inference performance. These results provide the fundamental insights for designing feedback mechanisms from realistic physical twins to virtual twins in the development of cardiovascular bio-digital twins.

\section*{References}
\bibliographystyle{IEEEtran}
\bibliography{references}

\end{document}